\documentclass[trackchanges,twocolumn]{aastex}

\usepackage[utf8]{inputenc}
\usepackage{natbib,epsfig,siunitx,url,graphicx,amsmath,footnote,float,xcolor,cases}
\begin{document}

\submitted{ApJL, accepted}

\title{Mercury as the relic of Earth and Venus' outward migration}

\author{Matthew S. Clement\altaffilmark{1}, Sean N. Raymond\altaffilmark{2} \& John E. Chambers\altaffilmark{1}}

\altaffiltext{1}{Earth and Planets Laboratory, Carnegie Institution for Science, 5241 Broad Branch Road, NW, Washington, DC 20015, USA}
\altaffiltext{2}{Laboratoire d'Astrophysique de Bordeaux, Univ. Bordeaux, CNRS, B18N, all{\'e} Geoffroy Saint-Hilaire, 33615 Pessac, France}
\altaffiltext{*}{corresponding author email: mclement@carnegiescience.edu}

\begin{abstract}

In spite of substantial advancements in simulating planet formation, the planet Mercury's diminutive mass, isolated orbit, and the absence of planets with shorter orbital periods in the solar system continue to befuddle numerical accretion models.  Recent studies have shown that, if massive embryos (or even giant planet cores) formed early in the innermost parts of the Sun's gaseous disk, they would have migrated outward.  This migration may have reshaped the surface density profile of terrestrial planet-forming material and generated conditions favorable to the formation of Mercury-like planets.  Here, we continue to develop this model with an updated suite of numerical simulations.  We favor a scenario where Earth and Venus' progenitor nuclei form closer to the Sun and subsequently sculpt the Mercury-forming region by migrating towards their modern orbits.  This rapid formation of $\sim$0.5 $M_{\oplus}$ cores at $\sim$0.1-0.5 au is consistent with modern high-resolution simulations of planetesimal accretion.  In successful realizations, Earth and Venus accrete mostly dry, Enstatite Chondrite-like material as they migrate; thus providing a simple explanation for the masses of all four terrestrial planets, inferred isotopic differences between Earth and Mars, and Mercury's isolated orbit.  Furthermore, our models predict that Venus' composition should be similar to the Earth's, and possibly derived from a larger fraction of dry material.  Conversely, Mercury analogs in our simulations attain a range of final compositions.

\end{abstract}

\section{Introduction}

While observations of star-forming regions indicate that the main ingredient for giant planet formation (free gas) dissipates within a few Myr \citep[though some observed disks are much older:][]{hernandez07}, isotopic analyses of terrestrial materials \citep[e.g.:][]{kleine09} suggest that the solar system's rocky worlds took shape over a more prolonged period of some tens to hundreds of Myr.  Recent cosmochemical studies have linked the respective planets' bulk compositions with those measured in analyses of chondritic meteorite groups presumed to represent fossilized remnants of the terrestrial system's building blocks.  Specifically, the Earth is thought to have accreted around 70$\%$ of its total mass \citep{dauphas17} from dry, highly-reduced Enstatite Chondrites \citep[EC,][]{javoy10}; the chemical constituencies of which also bear remarking similarities to those on the surface of Mercury as inferred via \textit{MESSENGER} X-ray spectroscopy \citep{nittler11}.  Conversely, analyses of the Martian meteorites indicate that Mars originated from an even mixture of EC material and Ordinary Chondrites \citep[OC,][]{tang14}.  In spite of these bulk similarities with non-carbonaceous chondrites (NC), D$/$H ratios of water on Earth closely resemble those in Carbonaceous Chondrites \citep[CC, e.g.:][]{daphaus00}.  This suggests the young Earth received a minor, and possibly late \citep{rubie15} contribution from wetter, CC material.

 In the classic theoretical picture of the solar system's genesis, the inner planets form within a circumstellar disk of $\sim$100 km planetesimals in the presence of the fully formed outer planets.  The terrestrial planets' intermediate precursors, often referred to as ``embryos,'' take shape by growing collisionally within this sea of asteroid-like objects \citep{koko_ida_96}.  This process in turn cultivates a bimodal size distribution of larger embryos and smaller planetesimals that continue to coalesce into the modern terrestrial planets through a series of giant impacts; one of the last of which is thought to have formed the moon \citep{wetherill78}.

Early numerical simulations of this scenario consistently formed analog systems comprised of $\sim$four, $\sim$Earth-mass terrestrial planets at roughly the proper semi-major axes within the correct amount of time \citep{chambers01,ray18_rev}.  As the modern masses of Mercury and Mars are only $\sim$6$\%$ and $\sim$11$\%$ that of the Earth, respectively, considerable focus over the past two decades of computational modeling has been placed on more accurately replicating the terrestrial system's precise mass distribution.  While a number of viable explanations for the Earth-Mars mass ratio have been proposed \citep[e.g.:][]{walsh11,ray17sci,clement18}, the Mercury-Venus ratio remains an extremely unlikely outcome in embryo accretion models \citep{ray18_rev}.  Other mysterious qualities of the solar system's innermost planet include its dynamical offset from Venus \citep[quantified by the ratio their orbital periods, $P_{V}/P_{M} =$ 2.6; much larger than those of the other neighboring terrestrial planets:][]{clement21_merc3}, large iron-rich core \citep[$\sim$70-80$\%$ of its total mass:][]{hauck13}, and the lack of additional planets inside of its orbit compared to those in exoplanet systems around Sun-like stars \citep[a plethora of which host short-period Super-Earths:][]{zhu18}.

In a series of recent papers, we identified a number of plausible terrestrial disk structures capable of more consistently replicating the Mercury-Venus mass and orbital period ratios, while also providing high-speed collisions that might remove mantle material from the young Mercury \citep[thus aiding in the reproduction its modern, iron-rich core:][]{benz88,asphaug14}.  In \citet{clement21_merc3}, we found that systems of 3-6 Mercury- to Mars-mass proto-planets in the vicinity of Mercury's modern orbit are easily destabilized by resonant interactions with Jupiter \citep{batygin15b}.  In successful simulations of this cataclysmic instability, excess planets merge with Venus or the Sun after undergoing a number of violent, erosive collisions; thus leaving Mercury behind as the sole survivor.  In a complimentary study \citep{clement21_merc2}, we investigated a scenario where local mass depletion in the region of embryos and planetesimals around Mercury \citep[e.g.:][]{lykawka17} is responsible for the planet's modern low mass and isolated orbit.  In particular, we found that the Mercury-Venus mass and orbital period ratios are best reproduced when the inner disk component extends from 0.35-0.75 au, possesses a total mass of $\sim$0.1-0.25 $M_{\oplus}$ that is dominated by planetesimals ($M_{pln}/M_{tot}=$ 0.5), and a shallow surface density profile that falls off with decreasing semi-major axis ($\Sigma \propto r^{0.5}$ for $r<0.75$ au).

While several authors have considered mechanisms that might produce a sharp, truncated inner edge in the terrestrial forming disk \citep[e.g.: a condensation-front or highly localized planetesimal formation:][]{draz16,morby16_ice}, such conditions do not regularly yield satisfactory Mercury analogs \citep{hansen09,ray18_rev}.  Though the initial conditions supposed in \citet{clement21_merc3} and \citet{clement21_merc2} might seem inconsistent with models of planetesimal formation and runaway growth, it is certainly plausible that the Mercury-forming region was dynamically reshaped after the onset of planetesimal formation.  \citet{ray16} proposed a mechanism for sculpting the surface density of embryos and planetesimals in the Mercury-forming region by hypothesizing that Jupiter's core formed in the innermost regions of the Sun's natal disk before migrating outwards to its modern location.  Indeed, modern hydrodynamical models of proto-planets evolving within nebular disks find that, depending on the thermodynamics of the disk, both inward and outward migration is possible \citep{kley08,bitch15} for proto-planets as small as a few tenths of an Earth mass \citep{bitch15}.  

Of particular interest to our present study, outward migration is typically favored at small radial distances near the end of the disk's lifetime due to strong negative temperature gradients.  These radial and mass dependencies of migration in turn give rise to ``convergence zones'' within the disk.  In these radial bins of null-migration, proto-planets of a particular size traversing the disk from inside-out stop and meet other objects migrating outside-in \citep[see][for an application of this concept to the solar system's terrestrial planets]{broz21}.

In this letter, we revisit the \citet{ray16} scenario, with the particular aim of understanding whether outward migration of embryos is a viable mechanism for reshaping the distribution of planetesimals within a terrestrial disk of more or less uniform surface density into one resembling the successful disk profiles identified in \citet{clement21_merc3} and \citet{clement21_merc2}.  To accomplish this, we leverage a suite of new high-resolution numerical simulations employing the GPU-accelerated (graphics process unit) code \textit{GENGA} \citep{genga}.  Our simulations include $\gtrsim$10,000 objects in the terrestrial region, analytic gas disk treatments, and forced outward migration through cubic interpolation of pre-determined orbital elements.  In addition to a scheme where Jupiter's migrating core restructures the terrestrial disk, we also simulate one where Earth and Venus' cores form in the vicinity of Mercury's modern orbit before migrating outward.  In this sense, our model is largely consistent with recent work \citep{broz21} successfully replicating the precise terrestrial system and favoring the planets' rapid accretion by invoking convergent migration (note that, in contrast to that study we do not consider material initially in the Mars-forming and asteroid belt regions, instead focusing our attention on the inner three terrestrial worlds).  Similarly, our work builds on studies considering the possibility of disk-wind-driven migration of planetesimals voiding the Mercury-region of planet-forming material \citep{ogihara18}.

\section{Methods}
\label{sect:meth_setup}

\begin{figure*}
	\centering
	\includegraphics[width=.99\textwidth]{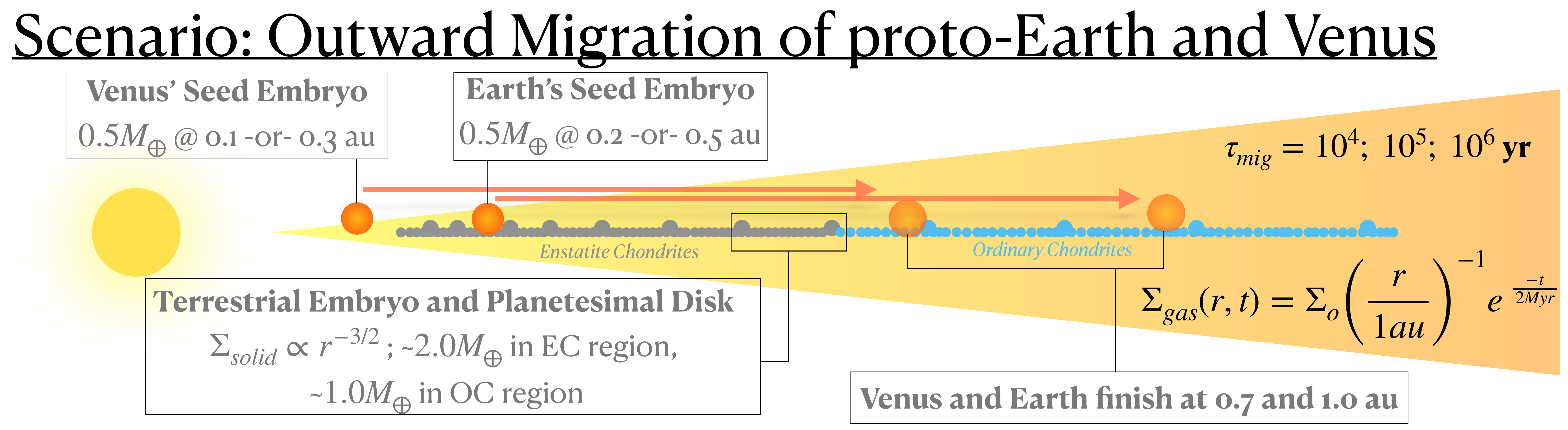}
	\caption{Graphical depiction of our modeled scenario.}
	\label{fig:cartoon}
\end{figure*}

Figure \ref{fig:cartoon} illustrates the general setup of our numerical models (see section \ref{sect:meth_sims} for specifics).  In all cases, we truncate the outer terrestrial disk at 1.1 au \citep[e.g.:][]{hansen09} as such initial conditions have previously been validated against a number of constraints \citep[e.g.:][]{walsh11,ray18_rev}, and enhance the probability of forming Mars analogs of the appropriate mass.  However, in principle our simulations should also be consistent with a scenario where a more massive terrestrial disk is truncated by the giant planet instability shortly after nebular gas dispersal \citep{clement18}.

In one collection of simulations (referred to as Jupiter-migration in the subsequent text), we revisit the scenario described in \citet{ray16} where Jupiter's 3.0 $M_{\oplus}$ core originates at the inner edge of the terrestrial disk.  In half of these models, Jupiter's core (and the terrestrial disk inner edge) is placed at 0.1 au, and in a second set of iterations we investigate an inner edge of 0.3 au.  The more extreme inner edge of 0.1 au not only maintains consistency with the work of \citet{ray16}, but is also motivated by the young Sun's presumed magnetic truncation radius \citep{frank92}, and empirically derived inner disk edges for planet forming regions with solar-like luminosities \citep{millan-gabet07}.  Conversely, our inner edge of 0.3 au is established to boost the probability of forming Mercury at the correct semi-major axis as described in \citet{clement21_merc2}.  The orbits of our embryos and planetesimals are determined in a manner consistent with most contemporary models of terrestrial planet formation \citep[e.g.:][]{chambers01,hansen09,clement18,ray18_rev}.  

In each of our initial models, we utilize 60 embryos and 10,000 planetesimals (the masses of the embryos are $\sim$10-100 times that of the planetesimals, depending on the simulation set) on nearly circular, co-planar, randomly oriented orbits with semi-major axes assigned such that the solid component of the disk's surface density profile is roughly proportional to $a^{-3/2}$ \citep[][Jupiter's core is considered in this calculation]{birnstiel12}.  Thus, our disks possess a total mass of material in the EC component (i.e.: the Mercury-forming region of $a<$ 0.75 au) of around 2.0 $M_{\oplus}$ (not including Jupiter's core), and 2.0 $M_{\oplus}$ in the OC section of the disk (0.75 $<a<$ 1.1 au; Earth/Venus-forming region).  All embryos and planetesimals besides Jupiter's core experience the gravitational potential of the gas disk, and migrate inward via Type-I migration if they grow large enough (see section \ref{sect:res_dyn} for a more detailed discussion of caveats related to this implementation).

In a second collection of simulations (Earth/Venus Migration) we suppose that proto-Venus and proto-Earth, rather than Jupiter's core, attain masses of 0.5 $M_{\oplus}$ within the EC region before migrating towards their modern orbits.  While it is unclear whether objects as massive as 3.0 $M_{\oplus}$ can form rapidly at small radial distances via pebble accretion \citep{boley14}, the expeditious formation of 0.5 $M_{\oplus}$ proto-planets inside of 0.5 au is consistent with modern models of planetesimal accretion during the gas disk phase \citep{morishma10,clement20_psj,woo21}.  We also structure these disks similar to those described above such that the total surface density profile is roughly proportional to $a^{-3/2}$ when proto-Venus and proto-Earth are included.  The remaining embryos and planetesimals total $\sim$1.5-2.0 $M_{\oplus}$ (depending on the inner disk edge) in the EC region and 1.0 $M_{\oplus}$ in the OC region.  

In principle, our models with an inner disk edge of 0.3 au our designed to recreate the successful disk profiles of \citep{clement21_merc2} after the outward migration phase is complete (i.e.: possess a mass deficit interior to 0.75 au and sufficient mass to form Venus, Earth and Mars in the OC region).  In addition to varying the inner disk edge, we test three migration timescales ($\tau_{mig}=$ $10^{4}$, $10^{5}$ and $10^{6}$, see section \ref{sect:meth_gas}) and two mass distributions in the EC region ($M_{emb}/M_{pln}=R=$ 1.0 or 4.0) that determine the precise masses of the embryos and planetesimals.  Thus, our study includes a total of 12 different models for each migration scenario (we perform two simulations for each model).  

\section{Results}
\label{sect:results}

\subsection{System Dynamics}
\label{sect:res_dyn}
We scrutinized the dynamical architectures of our modeled terrestrial systems against a number of common metrics \citep[e.g.:][]{chambers01}. In general, our simulations investigating the outward migration of Jupiter's core are far less successful at replicating the modern inner solar system than our models where Earth and Venus' seed embryos sculpt the Mercury-region.  Indeed, only one of our realizations where Jupiter migrates through the entire terrestrial disk produces reasonable analogs of Earth and Venus.  In the majority of cases, Jupiter accretes and scatters material from both the OC and EC regions of the terrestrial disk, thereby removing all but a few terrestrial embryos and leaving behind a distribution of planetesimals with insufficient total mass to form $\sim$Earth-mass planets.  As demonstrated in \citet{ray16}, this is occasionally avoided when Jupiter's migration is fast ($\tau_{mig}=$ 10$^{4}$ yr).  However, we find that in these cases embryos in the inner EC region of the disk are the most likely to avoid loss via accretion or gravitational scattering.  Therefore, the final planets formed in these quasi-successful simulations possess a radially dependent, hierarchical mass distribution.  This effect is not quite as severe when the planetesimal component of the EC region is more massive ($R=$ 1.0).  However, given the fact that our Earth/Venus-migration models outperform these cases at essentially every level of analysis, we focus the remainder of our discussion on Earth and Venus' outward migration.

\begin{figure}
	\centering
	\includegraphics[width=.5\textwidth]{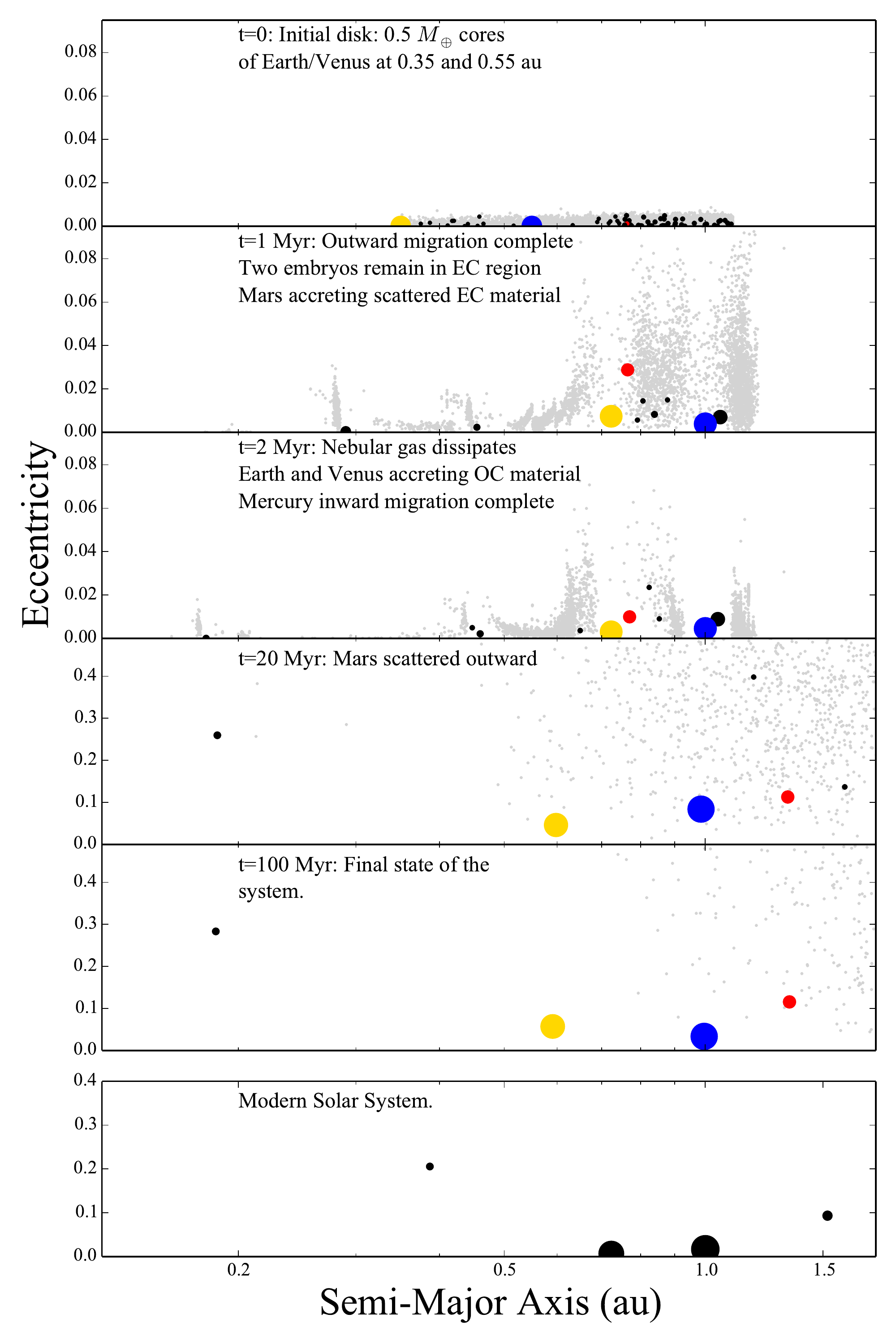}
	\caption{Example evolution of a simulation where Earth and Venus' 0.5 $M_{\oplus}$ cores originate at 0.3 and 0.5 au before migrating outward with a migration timescale, $\tau_{mig}=$ $10^{6}$ yr.  The semi-major axis and eccentricity of each object in the simulation is plotted at each time of interest, and the size of each point is proportional to the object's mass.  Embryos are color-coded black, planetesimals are plotted in grey, Earth is denoted with blue, Venus is indicated with gold, and Mars is colored red.  The final masses of the four planets are 0.055, 0.74, 0.92 and 0.19 $M_{\oplus}$, respectively.  We integrated this terrestrial system in the presence of the giant planets on their modern orbits for 500 Myr to verify its long-term dynamical stability.}
	\label{fig:time_lapse}
\end{figure}

In contrast to our Jupiter-migration scenario, we find slower migration ($\tau_{mig}=$ 10$^{6}$ yr and a minority of 10$^{5}$ yr runs) to be the preferable migratory route for proto-Venus and proto-Earth.  Indeed, the median fraction of EC region embryos and planetesimals lost after 2 Myr of simulation time in our $\tau_{mig}=$ 10$^{4}$ yr systems is 0.05.  Conversely, our $\tau_{mig}=$ 10$^{6}$ yr systems lose an average of 24$\%$ of the original EC mass via mergers with the Sun, ejection, or accretion by the migrating proto-Earth and Venus by $t=$ 2 Myr.  Figure \ref{fig:time_lapse} shows an example evolution where Earth and Venus' slow migration sculpts a primordial terrestrial disk extending from 0.3-1.1 au into a reasonable analog of the modern system.  As the two proto-planets migrate outward they accrete EC and OC material, while also initiating gravitational scattering events with embryos in the vicinity of Mercury's modern orbit that deplete the total mass in the inner region of the disk.  As Earth leads the pair in migration, it encounters a higher relative surface density of embryos and planetesimals than Venus.  In simulations where this is the case, Earth finishes the integration 24$\%$ more massive than Venus on average, thus explaining the modern mass of each planet.  An additional by-product of the planets' migration phase is the implantation of EC planetesimals into the asteroid belt \citep[e.g.:][]{ray17sci}.  We note that an average of 0.000376 $M_{\oplus}$ of EC material remains in the belt region after our 100 Myr simulations; thus providing a potential explanation for the origin of drier S-type asteroids similar to Vesta and extremely volatile-poor E-types \citep{zellner77}

If Earth and Venus migrated in tandem over an appreciable radial range, it is possible that they became trapped in an orbital resonance.  While such captures do not occur in our systems because our forced migration routes avoid resonance crossings, if this were the case, a subsequent dynamical exchange such as the Moon forming impact, a scattering interaction with Mars (panels 3 and 4 of figure \ref{fig:time_lapse}), or the giant planet instability \citep{bras13,clement18} would be required to dislodge the planets from resonance (the pair currently reside just outside of the 3:2 mean motion resonance).  It is also possible that the thermal state of the gas disk allowed the proto-planets to attain higher eccentricities during their migration, thus allowing them to avoid resonant capture \citep{broz21}.

Table \ref{table:mercs} tabulates several important average properties of the final Mercury analogs (all planets finishing with $a<$ 0.5 au) formed in our various simulation sets.  It is clear from figure \ref{fig:time_lapse} that the embryo that ultimately becomes Mercury undergoes substantial inward Type I migration.  As a result of this process, as well as Earth and Venus clearing the $a\lesssim$ 0.75 au region of additional planet forming material, Mercury attains an orbit that is extremely isolated from those of the other planets.  While the ratio between the Mercury and Venus analogs' orbital periods in this simulation ($\sim$5.0) is much more than that of the actual pair of planets (2.6), we argue that that this is due to the somewhat arbitrary length of time our simulations incorporate a Type I migration model.  As the end of Mercury's migration is related to the time of nebular dissipation, it would not traverse as far inward if, for example, nebular dissipation was quicker, or the planet did not grow large enough to begin to migrate substantially until some time had elapsed in the simulation.  

Given these caveats, and the fact that the final Mercury-Venus period ratios in our simulations are in the range of $\sim$2.5-6.0, we consider our disks with inner edges at 0.3 au to be highly successful at replicating Mercury's orbit.  Similarly, the median Mercury analog mass in our simulations is 0.051 $M_{\oplus}$ (Mercury's actual mass is 0.05 $M_{\oplus}$), and the average mass of all final planets with $a<$ 0.5 au is 0.15 $M_{\oplus}$ in our 0.3 au inner disk edge runs.  Thus, we conclude that Earth and Venus' outward migration is an efficient mechanism for producing a small Mercury analog.  We performed an additional 20, lower-resolution (1,000 planetesimals instead of 10,000) simulations of our preferred set of initial conditions ($\tau_{mig}=$ 10$^{6}$ yr; $a_{in}=$ 0.3 au; $R=$ 1.0, see the bottom column of table \ref{table:mercs}) for a first-order, statistical verification of their success.  Overall, these simulations yield excellent solar system analogs, and the results are consistent with those of our initial high resolution simulations in terms of the parameters reported in table \ref{table:mercs}.   Related to the above point on the Mercury-Venus period ratio, three of our 20 systems successfully form a single Mercury analog with 1.7 $<$ $P_{Ven}/P_{Merc}$ $<$ 2.7.

\begin{table*}
\centering
\begin{tabular}{c c c c c c}
\hline
Model & $N_{Merc}$ & $P_{Ven}/P_{Merc}$ & $e_{Merc}$ & $i_{Merc}$ & $M_{tot}$(a $<$ 0.5 au) \\
\hline
Jupiter-Migration; $\tau_{mig}=$ 10$^{4}$ yr & 5.5 & 2.5 & 0.018 & 0.58 & 1.42 \\
Jupiter-Migration; $\tau_{mig}=$ 10$^{5}$,10$^{6}$ yr & 6.0 & 2.4 & 0.0179 & 0.49 & 1.14 \\
Earth/Venus-Migration; $a_{in}=$ 0.1 au & 4.5 & 2.4 & 0.035 & 1.99 & 0.55 \\
Earth/Venus-Migration; $a_{in}=$ 0.3 au & 1.7 & 3.8 & 0.22 & 7.63 & 0.15 \\
20 simulation set & 1.5 & 3.3 & 0.14 & 5.4 & 0.11 \\
\hline
\textbf{Solar System} & 1.0 & 2.6 & 0.21 & 7.0 & 0.055 \\
\hline
\end{tabular}
\caption{Comparison of Mercury analog statistical properties in our various simulation sets with important qualities of the actual planet.  The columns are as follows: (1) The subset of simulations, (2) the average number of final planets (Mercury analogs) with $a<$ 0.5 au per simulation, (3) the orbital period ratio between the outermost Mercury analog and Venus, (4-5) the average eccentricity and inclination of the analogs and (6) the average total mass of all Mercury analogs at the end of the simulation.}
\label{table:mercs}
\end{table*}

In contrast to the system plotted in figure \ref{fig:time_lapse}, our models with disk edges at 0.1 au tend to produce systems of 3-4 $\sim$Mercury-mass planets with 0.1 $\lesssim$ a $\lesssim$ 0.5 au.  While these runs are not necessarily successful in the context of our current analysis, it is possible that they could be destabilized in a manner such that additional planets are lost as demonstrated in \citet{clement21_merc3}.  Indeed, resonant perturbations from Jupiter's eccentricity after it achieves its modern orbit and orientation with respect to Saturn \citep{nesvorny12,batygin15b} are essential for disrupting the orbits of objects in the Mercury-region.  As our simulations consider the giant planets on their pre-instability orbits, our multi-Mercury simulations tend to evolve on relatively stable orbits.  However, given the consistencies between the outcomes in this subset of our contemporary models and the initial conditions supposed in \citet{clement21_merc3}, we cannot rule out an inner disk edge of 0.1 au as inconsistent with Mercury's formation.

\subsection{Cosmochemical Implications}

\begin{figure}
\centering
	\includegraphics[width=.5\textwidth]{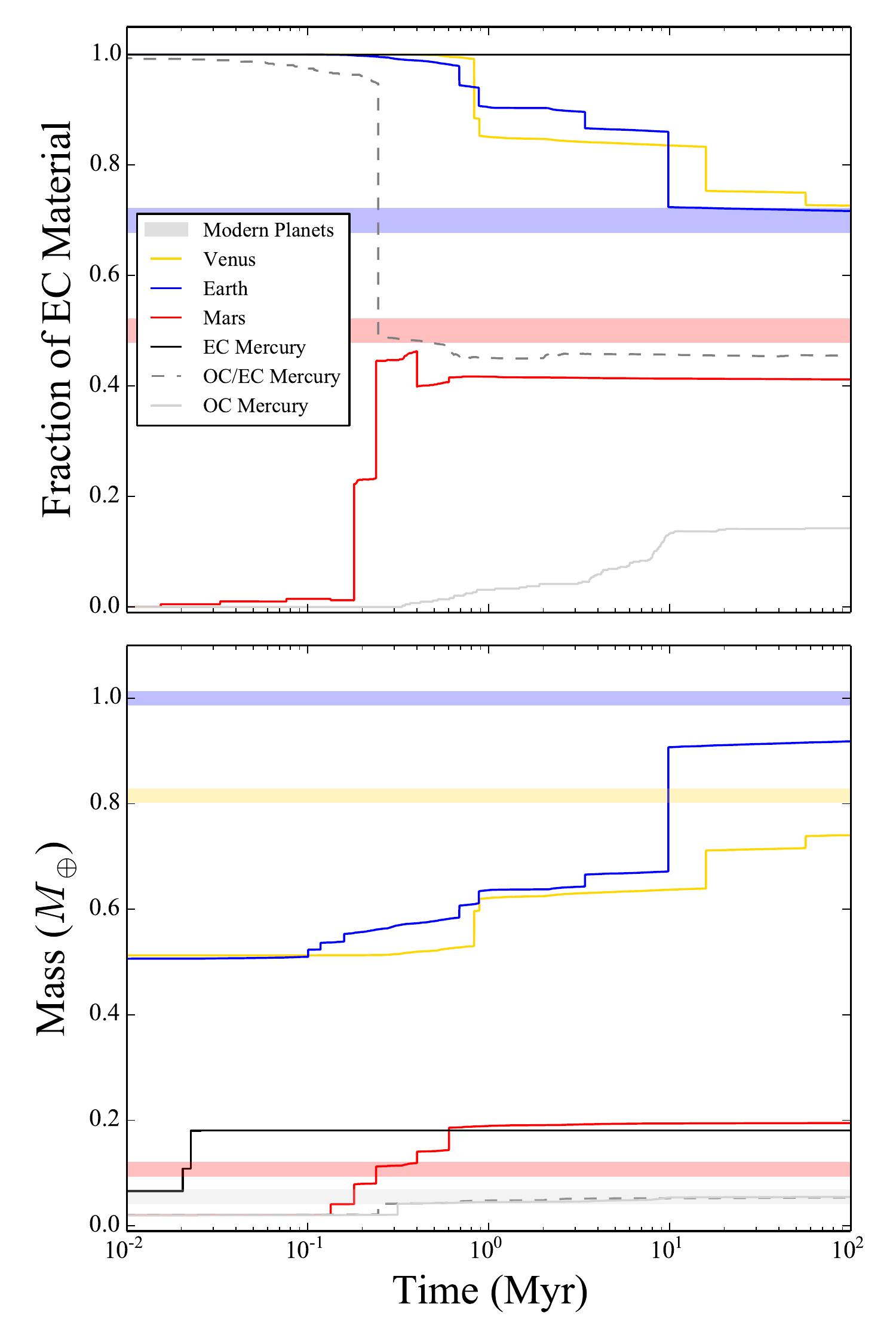}
	\caption{Cosmochemical implications of the outward migration scenario.  The top panel plots the compositional evolution of the four planets formed in the system depicted in figure \ref{fig:time_lapse} (black, gold, blue and red lines for the four respective terrestrial planets) assuming all material inside of 0.75 au is EC in origin, and embryos and planetesimals outside of 0.75 au have OC-like compositions.  The dashed and solid grey lines plot alternative evolutionary histories from different simulations where Mercury accretes mostly EC material or a mixture of EC and OC.  The horizontal grey regions plot the approximate chemically inferred contributions of each reservoir to the bulk compositions of Earth \citep[$\sim$70$\%$ EC and 30$\%$ OC, e.g.:][]{dauphas17} and Mars \citep[$\sim$50$\%$ EC and 50$\%$ OC, e.g.:][]{tang14}.   The bottom panel depicts the temporal evolution of the same planets' masses.  The four horizontal lines correspond to the modern masses of each respective planet.}
	\label{fig:feedzone}
\end{figure}

Figure \ref{fig:feedzone} plots the temporal evolution of the composition and mass of each planet from the simulation depicted in figure \ref{fig:time_lapse}.  In excellent agreement with the chemically inferred compositions of the respective planets \citep{javoy10,tang14}, the Earth analog in this run accretes 72$\%$ of its total mass from the EC region, while the Mars analog only draws 41$\%$ of its bulk material from the same region.  This provides a natural explanation for the ``snowline problem \citep[see][and references therein]{morby16_ice} by requiring Earth accrete the majority of its mass much closer to the sun before migrating outward.  Similarly, our model explains the Earth-Mars compositional dichotomy as the direct consequence of Earth first forming in the EC reservoir before migrating into a less collisionally evolved OC component of the disk.  This is also consistent with the heavy depletion of moderately volatile elements like Potassium \citep[e.g.:][]{allegre01} in the Earth's mantle, as well as models favoring an isotopically distinct, highly-oxidized Moon-forming impactor \citep[e.g.:][see the final impact on Earth in figure \ref{fig:feedzone}]{schonbaschler10}.  Such an origin for the Moon-forming impactor would require nearly complete homogenization \citep{nimmo10} of material in the aftermath of the collision to explain isotopic similarities between the two worlds \citep[for instance $\Delta ^{17}$O:][]{wiechert01}.  Though our simulations do not model the outer terrestrial disk in the vicinity of the modern asteroid belt, our scenario would necessarily imply that the majority of Earth's current volatiles were added via impacts with CC-type small bodies in the final phase of its accretion \citep[consistent with core-mantle differentiation models:][the former of which argued that accretion changed from mostly reduced to oxidized material after Earth reached $\sim$60$\%$ its present size]{rubie15,dauphas17}.

While our presumed initial EC/OC divide is admittedly contrived (see section \ref{sect:limits}), our simulations investigating Earth and Venus' outward migration consistently produce highly disparate accretion histories for Earth and Mars (both in terms of radial feeding zone and growth rate).  Venus analogs tends to draw quantities of material from the EC and OC regions that are similar to those accreted by Earth analogs, and often finishes with a slightly drier composition than Earth provided it does not suffer a significant late collision with a massive embryo from the OC region \citep[note that Venus' lack of a natural satellite and magnetic field might evidence it having avoided such a late collision analogous to the Moon-forming impact:][]{jacobson17b}.

While the final giant impact on Earth in the simulation depicted in figures \ref{fig:time_lapse} and \ref{fig:feedzone} occurs around $t=$ 10 Myr \citep[much quicker than the $\sim$30-60 Myr timing of the Moon-forming impact as inferred from Hf-W chronology:][]{kleine09}, this is largely attributable to the stochasticity of the final giant impact timing in our simulations.  Indeed, other simulations produce Moon-formation events at later epochs.  For instance, one system in our additional suite of 20 low resolution simulations produces a final giant impact at $t=$ 52 Myr with a projectile to proto-Earth mass ratio of 0.19 (the average final giant impact time in these simulations is 7.62 Myr).  Thus, in spite of the fact that the majority of Earth's accretion occurs in the first few Myr of our simulations, our scenario does not conflict with the planet's growth chronology as inferred via isotopic dating.  Moreover, the precise timing of the Moon-forming impact is still a matter of debate.  Depending on the depth and degree of metal-silicate equilibration after the event, impacts as early as 10 Myr are potentially consistent with Hf-W analyses \citep{fischer17}

Our Mars analogs tend to exhibit extremely rapid accretion that is mostly complete within the life of the nebular disk.  Consistent with past studies that also truncate the terrestrial disk around Earth's modern orbit \citep{hansen09,walsh11}, Mars' growth is stunted in the majority of our models when it is scattered out of the central component of the disk via gravitational encounters with Earth or Venus.  While this is consistent with Hf-W dating of the Martian meteorites \citep{Dauphas11}, it is also possible that Mars' growth was halted by an early giant planet instability \citep{clement18}.

Our simulations are far less deterministic in regards to Mercury's compositional origin.  Figure \ref{fig:feedzone} depicts three types of evolutionary paths followed by our Mercury analogs.  If Mercury originates in the OC region, it typically does not migrate into the EC region until after Earth and Venus remove the majority of the reservoir's massive embryos.  Thus, it can survive the formation epoch with a mostly-OC composition (solid grey line).  Conversely, if Mercury originates in the EC component and happens to avoid late impacts with scattered OC embryos, its ultimate composition is nearly 100$\%$ EC (black line).  When this is not the case, however, the planet can acquire a significant fraction of its mass from OC material in spite of originating in the innermost section of the disk (dashed grey line).  In most cases though we note that the final planetesimals that merge with Mercury (i.e.: its late veneer) originate in the EC component.  This result is broadly consistent with \textit{MESSENGER} observations of the planet's surface \citep{nittler11}.

Efforts to infer the relative timeline of EC and OC incorporation into the young Earth are complicated by several lines of contradictory evidence.  Analytic metal–silicate partitioning models considering a number of lithophile and moderately siderophile elements (namely $\Delta^{17}$O, $\epsilon^{48}$Ca, $\epsilon^{50}$Ti, $\epsilon^{54}$Cr, $\epsilon^{50}$Ni, $\epsilon^{92}$Mo and $\epsilon^{100}$Ru in \citet{dauphas17} favored the first 60$\%$ of Earth's accretion involving an even mixture of EC and OC-type material before transitioning to an entirely EC phase of growth.  The OC material in stage one is required to reproduce slight dissimilarities between the Earth and E chondrites in Calcium and Titanium.  The second phase of 100$\%$ EC growth is required to reproduce the Molybdenum signature in Earth that is inconsistent with an OC origin.  Although this proposed accretion history is in conflict with our proposed scenario (figure \ref{fig:feedzone}), alternative models \citep[e.g.:][]{schonbaschler10} support the late accretion of wetter material.  Therefore, we conclude that certain lines of cosmochemical evidence are inconsistent with our model, while others might favor it.

\section{Conclusions}

In this letter we presented a collection of numerical simulations designed to investigate a scenario where outward migrating proto-planets reshape the terrestrial disk into a structure that favors the formation of a small Mercury.  We favor a scheme where 0.5 $M_{\oplus}$ seeds of Earth and Venus form rapidly within the nebular disk around 0.3 and 0.5 au as demonstrated in recent high-resolution simulations of planetesimal accretion \citep{clement20_psj,woo21}.  With the precise thermodynamic structure of the Sun's natal disk still largely unconstrained, the validity of our scenario relies heavily on an assumption that the terrestrial forming region attained a thermal profile that supported the outward migration of $\sim$Earth-mass bodies around the time Earth and Venus attained half their modern masses.  If this were the case, the planets' outward migration provides a natural explanation for several orbital and bulk chemical properties of the planets.  The Mercury-Venus mass ratio, as well as the dynamical isolation of the innermost planets' orbit, are attributable to Venus and Earth accreting and scattering material in the vicinity of Mercury's current orbit as they traverse the disk.  Similarly, the Earth-Venus mass ratio is a consequence of Earth migrating ahead of Venus.  Through this epoch of migration, both Earth and Mars accrete respective mixtures of EC and OC-type material that are consistent with each planets' inferred bulk composition \citep{javoy10,tang14}.  Moreover, our model predicts that Venus' chemical makeup should be nearly identical to the Earth's.  In some cases, Mercury forms completely out of dry EC material.  While this is consistent with \textit{MESSENGER} spectroscopy, we also note instances where Mercury originates in the outer, presumably OC component of the disk, and others where the planet forms out of a mixture of both reservoirs.  Future work must statistically validate our proposed scenario with a larger collection of simulations.  

\section*{Acknowledgments}
The authors thank Conel Alexander, Alan Jackson and an anonymous reviewer for useful discussions and input that greatly improved the quality of this manuscript.  S.N.R. acknowledges support from the CNRS’s PNP program.  This work used the Extreme Science and Engineering Discovery Environment (XSEDE), which is supported by National Science Foundation grant number ACI-1548562. Specifically, it used the Bridges2 system, which is supported by NSF award number ACI-1445606, at the Pittsburgh Supercomputing Center \citep[PSC:][]{xsede}.
\bibliographystyle{apj}
\newcommand{\sci}{$Science$ }
\bibliography{merc4.bib}

\appendix
\section{Supplementary Information}
\subsection{Numerical Simulations}
\label{sect:meth_sims}

Each of our numerical simulations utilize the GPU-accelerated, $GENGA$ integration package \citep{genga} that is optimized to run on all \textit{Nvidia} GPUs.  With the exception of our artificial gas disk treatments (section \ref{sect:meth_gas}), we utilize settings that are standard in N-body studies of terrestrial planet formation.  Objects that make perihelia passages less than 0.025 au are merged with the Sun, those exceeding 20 au at aphelion are considered ejected, and colliding objects are merged without fragmenation.  Simulations where a migrating core begins at 0.1 au use a timestep of 0.5 days, and those truncating the EC region at 0.3 au use 3.65 days.  All planetesimals are treated as semi-active particles by the code; meaning that they feel the gravitational forces of the larger embryos, but do not perturb or collide with one another.  Embryos experience the gravity of, and can collide with all objects in the simulation.  Each simulation is run for a total integration time of 100 Myr.  At $t=$ 2 Myr, we remove all artificial forces, and incorporate modern-massed versions of Jupiter and Saturn in a 3:2 mean motion resonance \citep[e.g.:][]{nesvorny12}.  This initial resonant configuration is consistent with hydrodynamical studies of the gas giants' early orbital evolution \citep[e.g.:][]{morbidelli07}, and axiomatically implies that they must attain their modern orbital configuration through an episode of dynamical instability after gas disk dissipation.  While a loose consensus has developed in recent years in favor of the solar system's instability occurring within the first 100 Myr after nebular dispersal, if the event transpired much earlier \citep[$t\lesssim$ 10 Myr:][]{clement18} it likely also affected Mars' formation.  Regardless of the specific timing, a resonant Jupiter and Saturn is the most plausible orbital configuration for our models of the gas disk phase.

\subsection{Gas Disk Model}
\label{sect:meth_gas}

Inward and outward migration regimes are strongly tied to the local thermodynamical properties of gaseous nebulas, which are intrinsically challenging to constrain when considering the Sun's primordial disk \citep[e.g.:][]{kley08,lega15,bitch15}.  Therefore, we opt to follow the example of \citet{ray16} and explicitly presume that our simulations begin at a point where conditions in disk support outward migration for proto-planets larger than $\sim$0.5 $M_{\oplus}$ \citep[see, for example, figures 7 and 9 of][]{bitch15}.  Thus, the validity of our modeled scenarios is tied to two key assumptions: (1) 0.5-3.0 $M_{\oplus}$ objects were able to form rapidly in the inner regions of the Sun's primordial gas disk and (2) the disk subsequently attained a thermal profile that favored the outward migration of these large objects between their formation locations and modern semi-major axes.  While previous modeling work provides justification for the plausibility of both assumptions, we argue in section \ref{sect:results} that the ability of our models to match multiple dynamical and cosmochemical constraints for the inner planets strengthens the case for our scenario.
 
 To more rigorously control the migration patterns of Jupiter, Earth and Venus, we utilize \textit{GENGA}'s built-in cubic interpolation algorithm to artificially force the protoplanets' outward radial migration along a pre-determined path sampled at 1 year intervals.  As described in \citet{ray16}, we assign migration sequences that are related to the radially dependent isothermal timescale for Type I migration \citep{tanaka02,paardekooper11}: \begin{equation}
 	\tau_{mig} = \bigg( \frac{1}{2.7 + 1.1\alpha} \bigg) \bigg( \frac{M_{*}}{m} \bigg) \bigg( \frac{M_{*}}{\Sigma a^{2}} \bigg) \bigg( \frac{H}{r} \bigg)^{2} \Omega^{-1}
\label{eqn:tau_mig}
\end{equation}  Here, $\alpha$ (set to 1.0 in our simulations) is the disk's surface density profile given by $\Sigma_{gas} \propto r^{-\alpha}$ (figure \ref{fig:cartoon}), $\Omega$ is the local Keplerian frequency and $H/r$ is the disk's vertical scale height (we use a flared profile $\propto a^{1/4}$ with $H/r=$ 0.034 at 1 au).  Given our chosen disk parameters, $\tau_{mig} \simeq 10^{4}$ years for a 3.0 $M_{\oplus}$ core initialized at 0.1 au, and around an order of magnitude larger for a 0.5 $M_{\oplus}$ core.  However, in light of the significant uncertainties involved with estimating outward migration timescales \citep[e.g.:][]{lega15}, we opt to test values of $\tau_{mig}=$ $10^{4}$, $10^{5}$ and $10^{6}$ yr for each of our modeled scenarios (section \ref{sect:meth_setup}).  Throughout the outward migration phase, the proto-planets' eccentricities and inclinations are forced to remain close to zero.

The remaining embryos and planetesimals in our simulations' orbits are perturbed by the gas disk using a standard implementation \citep{morishma10} that is commonly employed in N-body simulations of planetesimal accretion \citep[e.g.:][]{clement20_psj,woo21}.  Specifically, we utilize $GENGA$'s built-in gas disk model that leverages artificial forces to model the effects of aerodynamic drag, tidal interactions that induce eccentricity and inclination damping \citep{paardekooper11}, and the nebula's global potential.  Thus, embryos that grow large enough (besides the outward migrating proto-planets) drift inward via Type-I migration, however we also experimented with models that did not incorporate aerodynamic drag forces.  In our model, the gas density decays uniformly in space and exponentially in time (figure \ref{fig:cartoon}).  For all of our simulations we utilize a gas dissipation timescale of $\tau_{gas}=$ 2 Myr that is consistent with the presumed lifespan of the Sun's gaseous disk \citep[note that some observed gas disks are much older, see][and references therein]{hernandez07,morishma10,clement20_psj}.

\subsection{Model Limitations}
\label{sect:limits}

Our present investigation should serve as a proof-of-concept of the outward migration scenario, rather than a robust rebuke of decades of terrestrial planet formation literature.  Notably, our models do not incorporate material in the asteroid belt \citep[an important constraint for terrestrial planet formation models:][]{walsh11,ray17sci}, neglect the question of water delivery to Earth, do not model the giant planets' growth phase, only test a single gas disk model, and decouple the vast majority of accretion in the inner solar system from the epoch of giant planet migration and instability \citep{clement18}.  In particular, the giant planet instability likely affected Mercury's orbit \citep{bras13}, as well as those of the other terrestrial planets \citep{bras13,clement18}.  Thus, future work must further develop this scenario within the broader picture of the young solar system's global evolution, and incorporate full hydrodynamical simulations to more concretely understanding the range of plausible migration paths \citep[e.g.:][]{broz21}.  

The implementation of our gas disk model is another notable limitation of our current study.  Rather than deriving the migration regimes of our proto-planets directly from a set of supposed disk parameters, we explicitly assume that the Sun's disk supported Earth and Venus' outward migration, and then test a range of plausible migration paths.  In this sense, our work strives to infer certain qualities of the primordial nebula (i.e.: a strong negative temperature profile late in its life) from the inner planets' modern dynamical configuration.  While this methodology is useful given the unconstrained nature of the Sun's natal disk, it also neglects several important physical processes that might affect the migration of objects in our simulations.  In particular, Lindblad and co-rotation torques \citep[driven by interactions with resonant density waves and gas traversing the disk on horseshoe-like orbits, respectively:][] {cresswell08,paardekooper11} depend on the eccentricity of the migrating proto-planet.  The average eccentricity of proto-Earth and Venus in our simulations during their migration phase is $\sim$0.005.  In this $e<2h$ regime, the Lindblad torque produces inward migration, and is thus inconsistent with our simulated migratory direction.  However, other phenomena such as the heating torque \citep{bentez15} and hot-trail effect \citep{chrenko17,eklund17} affect the migration direction and eccentricities of planetary cores and embryos, and can potentially jointly conspire to support outward migration in certain regions of the terrestrial disk.  For a more detailed exploration of outward migration in the inner solar system considering each of these effects, we direct the reader to \citet{broz21}.

In contrast to some recent studies of terrestrial planets' embryo precursors' evolution within the primordial nebular gas \citep[e.g.:][]{levison15,chambers16,ormel17,lambrechts19}, our simulations do not incorporate a pebble accretion model.  While this simplification potentially affects the growth histories and final masses of the analog planets generated in our models, bulk isotopic differences between the main NC and CC meteorite groups \citep[e.g.:][]{budde16} have been interpreted to imply that two distinct reservoirs of planet-forming material were spatially separated early in the solar system's history.  Under a strict interpretation of this constraint, inward drifting pebbles from the outer disk could not have contaminated the compositions of inner solar system planetesimals around the end of the gas disk phase \citep{kruijer17}. A similar assumption in our simulations is related to the initial location of the EC/OC divide.  While planetesimal formation models \citep[e.g.:][]{draz16} will be required to validate these initial conditions, the more important result of our study is that Earth and Mars are consistently built from different regions of material.

\end{document}